# Blockchain for Decentralized Multi-Drone to Combat COVID-19

S. H. Alsamhi[1,2], B. Lee[1], M. Guizani3, N. Kumar[4], Y. Qiao[1], Xuan Liu[5]
[1]Software Research Institute, Athlone Institute of Technology, Athlone, Ireland.
[2]IBB University, Ibb, Yemen
[3]Qater University, Qater
[4]Thapa Thapar Institute of Engineering and Technology, Patiala (Punjab), India
[5]College of Information Engineering (College of Artificial Intelligence), Yangzhou University, China

**Abstract: -**Currently, drones represent a promising technology for combating Coronavirus disease 2019 (COVID-19) due to the transport of goods, medical supplies to a given target location in the quarantine areas experiencing an epidemic outbreak. Drone missions will increasingly rely on drone collaboration, which requires the drones to reduce communication complexity and be controlled in a decentralized fashion. Blockchain technology becomes a must in industrial applications because it provides decentralized data, accessibility, immutability, and irreversibility. Therefore, Blockchain makes data public for all drones and enables drones to log information concerning world states, time, location, resources, delivery data, and drone relation to all neighbors drones. This paper introduces decentralized independent multi-drones to accomplish the task collaboratively. Improving blockchain with a consensus algorithm can improve network partitioning and scalability in order to combat COVID-19. The multi-drones task is to combat COVID-19 via monitoring and detecting, social distancing, sanitization, data analysis, delivering goods and medical supplies, and announcement while avoiding collisions with one another. We discuss End to End (E2E) delivery application of combination blockchain and multi-drone in combating COVID-19 and beyond future pandemics. Furthermore, the challenges and opportunities of our proposed framework are highlighted.

## I. Introduction

The COVID-19 outbreak is to fully quantify the value of advanced digital technologies to pandemics response in the early stage. Digital technologies represented the critical technology to support smart healthcare response to COVID-19 outbreak worldwide, such as identifying cases, monitoring the population, delivering goods and medical supply, decentralized alert, zero touches remote monitoring, and detecting cases from faraway. Therefore, smart healthcare for responding to COVID-19 and future pandemics is become more digital [1]. The advanced technologies are used to mitigate the impact of the COVID-19 outbreak, such as Artificial Intelligence (AI), the Internet of Things (IoT), drone, robotics, 5G, blockchain [2]. Furthermore, the authors addressed the numerous advantages that drones and blockchain could provide to mitigate the impact of COVID-19. These advanced technologies accelerate faster than before to create a safer environment world. Thus, the opportunities for research areas are discussed the combining several technologies collaboratively towards combating the COVID-19 outbreak [3]. However, the combination of advanced and emerging technologies can improve the ways to combat the COVID-19 outbreak and future pandemics such as multi-robot collaboration and blockchain technology [4], AI and drone technology [5], AI for robots [6, 7], blockchain, and AI [8]. In [9], the collaboration among multi-user concept is discussed and how they perform common task efficiently with recognizing the activity.

Drones are flying in the low altitude platform [10-12]. Using multi-drone could deliver goods and serve people in the quarantine zone to fulfill their needs and purposes. However, a drone network managing swarm needs efficient techniques and protocols to perform the required tasks efficiently in the quarantine area. Furthermore, a swarm of drones needs collaboration to share locations, trajectories, tasks, and purpose. With the support of the IoT, AI, and Cloud/Edge computing, multi-drone is enabling complex interactions amongst themselves. Multi-drone interacts with each other in order to exchange information, interdependent subtask, share goals, mutual trust, mutual learning, and mutual adaption [4]. However, the existing challenges that impede large-scale multi-drone (e.g., swarm robotics [7]) include drone network architecture, drone supervision, network partitioning, scalability, time, safety, and energy efficiency. The above issues can be solved by blockchain technology.

In drone network architecture, centralized suffers from a single drone failure, while decentralized suffers lack all information about drones in the network. In the case of centralization, decision making will take a long time to control a swarm of drones during their task performances, and therefore, the collision between drones occurs because of response delay. Therefore, decentralized can improve drone network performance and enhance Quality of Service

(QoS) and reduce task performing time. Blockchain technology is also decentralized and adapted to quickly ensure many actions during the networks drone interaction [13]. Moreover, blockchain can improve the speed and how swarm drones change their behavior during task performance due to global information for addressing specific drone requirements. Furthermore, a swarm drones system needs to accommodate many drones and simultaneously maintain the required level of robustness. The multi-drone can be partitioned dynamically due to noise in the channel communication between partitions. The system needs to complete the tasks within a given time or energy limits. However, blockchain improves the swarm drones system's global data, resulting in high productivity and sustainable maintenance.

To address challenges, blockchain has proposed in many works to facilitate drone collaboration efficiently. By using blockchain, independent drones reach consensus in a decentralized fashion. A blockchain serves as a shared knowledge to improve the performance of the system [14]. Smart contracts support great potential in order to enable a high level of security [15], operations of robotic [16], flexible, autonomous, and profitable [17]. The current work is still at an early stage. This conceptual framework paper discusses developing a decentralized ledger platform with required protocols to enable multi-drone collaboration networks with malfunctioning drones and network partition tolerance. The focus is to verify the feasibility of applying blockchain technology in a multi-drone to perform complex tasks in the quarantine area. Most solutions require the drone to process and store the current status transactions equally and collaboratively, just as blockchain technology does. We address the above challenges by designing a novel consensus algorithm for multi-drone collaboration to combating the COVID-19 outbreak efficiently, focusing on network partitioning, scalability, and byzantine drones.

### A. Contribution

It is believed that blockchain can help in the decentralized multi-drone for combating COVID-19 and the impact on pandemic management situations. To date, none of the literature address blockchain for multi-drone decentralized on fighting COVID-19. The primary purpose of this conceptual framework paper is to provide the readers with a full picture of how blockchain technology powered multi-drone decentralized for delivering services in the quarantined area during the outbreak of pandemics. Therefore, in this conceptual framework paper, we present the challenges and different proposed frameworks that motivate other researchers to work extensively on combating COVID-19 pandemics. To this end, our conceptual framework paper focuses on the following contributions:

1) We present a conceptual framework architecture that combines blockchain and multi-drone for combating COVID-19, aiming to find efficient solutions in response to COVID-19 and future pandemics.
2) We highlight the blockchain potentials in decentralized multi-drone collaboration and discuss its applications for combating COVID-19 and future pandemics.
3) We discuss two use cases for the multi-drone to combat COVID-19, i.e., monitoring and detecting COVID-19 and E2E delivery systems to people in the quarantined area.
4) Based on the conceptual framework paper, we provide the research challenges, opportunities, and different proposed frameworks to motivate and encourage other researchers to work extensively on combating COVID-19 pandemics.

### B. Paper scope and organization

This conceptual framework paper focused mainly on controlling and managing multi-drone collaboration to combat the COVID-19 outbreak and future pandemics with blockchain technology. The rest of the paper is organized as a fellow. Related work is discussed in Section II, while the component and system architecture described in section III. Section IV then describes the proposed framework solution and application and discussion described in Section V and VI, respectively. Finally, the work conclusion concludes in section VII.

## II. Related work

Drones are used for many purposes to combating COVID-19, such as vigilance, monitoring, food delivery, medicine delivery, thermal scan, ensuring social distance, sanitization, and alert system [18, 19]. The authors of [20] discussed how to gather data in real-time from wearable sensors, thermal scanners, cameras, and movement sensors. They considered drone as an edge intelligent in order to process data gathered and to avoid collision.

The multi-drone has comprehensive application scenarios including unreached, unknown, and dangerous environments (e.g., disaster Search and Rescue teams (SAR) [21, 22], humanitarian demining, underwater exploration, or surveillance [15], quarantine area [23]. The consensus issue among multi-drones is essential to monitor, detect, and deliver to improve real-time decision-making [24]. In many scenarios, centralized control is an issue because drones or robots have to communicate with each other in a peer-to-peer manner. Drones may need to adapt to different emergent situations. Some drones may be hacked, use faulty sensors or noisy communication channels, and lose availability. An emerging trend in this community is to use blockchain technologies to tackle these challenges. In [25], the authors presented the advantages of blockchain, and multi-robot combination, including security, management and control, robot behavior differentiation and distributed decision making.

In [15], the authors proposed blockchain for decision making a collection in the swarm robotics. They compared the blockchain approach with a probabilistic finite state machine (PFSM) [26]. The findings showed that PFSM breaks down even if a small number of byzantine robots, while blockchain provides a robust solution. In [27], the authors implemented a distributed ledger for multi-robots using BigchainDB to manage the feasibility. The distributed ledger was including Parrot Jumping Night ground drones and Parrot AR Drone 2.0. In [28], the authors proposed a swarm Directed Acyclic Graph (DAG) called swarmDAG for multi-robot partition tolerant of network. SwarmDAG enabled multi-robot to achieve consensus.

The growth of blockchain technology is exponentially in the robotics domain because it allows robots to conduct a transaction without third party authority. Hence, blockchain technology becomes a must in industrial applications due to its advantages to support decentralized data, immutability, accessibility, non-repudiation features, and irreversibility. Recently, the advantages make blockchains the probably and promising technology in smart industries [29]. The authors of [30] presented blockchain for controlling and coordinating multi-agent consisting of drones. For example, drones frequently interact with others nearby. This "community" feature [28] can be explored in designing the consensus algorithm, e.g., the consensus in the neighborhood is probably valid in the global scope because their transactions (interactions) are not relevant to the other communities. According to the network physics system's environment, the authors of [20] introduced a dynamic distributed process control model for attack defense resources. Blockchain technologies will be evaluated as the best way for decentralization and collaboration of participants in many applications.

The computational resources represent the scalability challenges of PoW [31]. On the other hand, the scalability challenges of PBFT is due to broadcasting messages and the cost of communication [32]. The authors of [33] introduced the network partitioned in subnetworks with improving blockchain scalability. Elastico implemented the sharding technique for the permissionless blockchain. Then, the authors of [34] proposed Merkle Patricia trees or Merklix trees in which it was used a different shard to global state [35]. However, scalability is the main challenge of blockchain technology. To solve the scalability issue, the sharding technique can play a vital role in enhancing blockchain scalability. The sharding architecture matches with the features of the community. Sharding techniques can enhance fault tolerance. Off-chain channels can be employed as complementary to sharding.

The work in joining blockchain and drones also covers other perspectives. Robotchain [36] introduces the record of robot action using a private Tezos blockchain. Then, the events recorded are going to be deciding the robot required for action performance in monitoring, conflicting situations, and tuning and behavior of the robot performance. Furthermore, DezCom model [37], proposes decentralized smart industry 4.0 applications focusing on multi-robot collaboration. The proposed DezCom implemented on Robot Operation System (ROS) with the Tendermint consensus algorithms help.

For securing data acquisition, the authors of [38-40] used blockchain technology for securing data gathered from IoT devices environment by drone technology. In [38], the findings showed that the proposed technologies' performance improves security, connectivity, and energy consumption. While [39] used blockchain technology for securing healthcare data gathered from IoT devices by drone. Blockchain technology, AI are applied for securing the drones collaboration [41], while the optimal techniques are used to ensure multi-drone collaboration task allocation with low energy consumption in a high level of security [42] and enable trustworthy communication among drones [43].

Furthermore, in [40] the authors introduced blockchain technology for processing data gathered by drones from IoT environments. Simultaneously, [30] authors predicted the signal strength from drones over IoT framework to gather data and improve energy efficiency. Also, the authors of [44] used blockchain for drone-enabled mobile edge computing. The smart contract was used to access the public, while the sub blockchain was used to improve stability. Recently, the authors of [45] introduced blockchain technology that envisioned softwarized for the communication of multi-drone-based 6G to tackle COVID-19. The results showed better performance in QoS as compared to 4G and 5G. QoS parameters and techniques for improving communication via space technologies discussed [46-51]. The authors of [52] introduced the collaboration of drones and Hetnet for enhancing QoS. The authors of [4, 8, 53] discussed the combination technologies for combating COVID-19, such as AI, blockchain, and robotics to combat COVID-19 epidemics. In contrast, the authors [4] discussed the combination of blockchain and multi-robot collaboration to combating COVID-19 via E2E delivery, detection, and identification and monitoring.

Table.1 comparative analysis of existing work and current work for combating COVID-19

| Ref. | Highlighted | Applications | | | Features | | | | | Technology | |
|------|-------------|--------------|---|---|----------|---|---|---|---|------------|---|
| | | COVID-19 | Disease | Disaster | Monitoring | Data gathered | Thermal image | Face detection | Delivery | Blockchain | AI |
| [54](2020) | Drone technology for support patients healthcare services | | √ | | √ | √ | | | √ | | |
| [11](2019) | Drone support internet of public safety things | | | √ | √ | √ | | | √ | | |
| [55](2020) | Drone technology for medical services | | | | √ | √ | | | | | |
| [45](2020) | Multi-drone tackle COVID-19 | √ | √ | | √ | √ | | | √ | | |
| [56](2018) | Drone technology for vacant parking detection | | | | √ | √ | | | | | |
| [57](2020) | Blockchain for securing drones delivery | √ | √ | | √ | | | | √ | √ | |
| [8](2020) | Technologies to combat COVID-19 | √ | √ | | √ | | | | √ | √ | √ |
| [58](2019) | 5G and drone technology for healthcare | | | | √ | | | | | | |
| [39](2020) | Blockchain for securing data gathered from IoT by drone | | | | | √ | | | | | |
| [59](2020) | Attack in healthcare device and drone technology | | | | √ | | √ | | | | |
| [4](2020) | Collaboration of multi-robot to fight the COVID-19 outbreak | √ | √ | | √ | √ | √ | √ | √ | √ | |
| [2](2020) | Role of advanced technologies and impact in COVID-19 | √ | √ | | √ | √ | √ | √ | √ | | √ |
| [20](2020) | Drone and techniques to combat COVID-19 | √ | √ | | √ | √ | √ | √ | √ | | √ |
| Work | Blockchain for decentralized multi-drone collaboration to fight COVID-19 and future pandemics | √ | √ | √ | √ | √ | √ | √ | √ | √ | √ |

**III. Component of system architecture**

The drone can deliver goods and medical supplies in real-time. Therefore, it could improve the emergency response to combat disease outbreak such as Corona, Lassa, Ebola, Malaria (detect, take and catch blood samples of mosquitos for better mitigate the potential of infectious disease outbreaks) [60]; medicine transportation (transport patient samples from affected people and a local disease center due to quarantine restrictions); and delivery goods (carrying samples and dispatching them for the test in case the quarantined restricted at their homes). It is highly recommended to ensure adequate disinfection and protection, usually gathering patients and staff to deliver goods to receive supplies without risking the spread of infection. Therefore, drones help scientists to respond and understand the way diseases become epidemics. For instance, China government mounted drones with loudspeakers to guide and deliver information for Wuhan's people [61]. Moreover, drones with thermal cameras are used to monitor the body temperature so that medical staff can quickly identify the affected new cases. Drones function for combating COVID-19 is summarized, as shown in Fig.1.

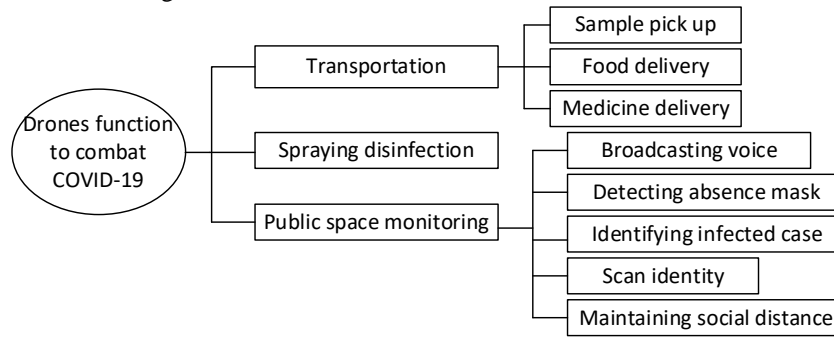

Fig.1 Drones function to combat COVID-19

Robotics systems are typically designed for running over long periods, which may accumulate a large size of blockchain and the states, making it impractical for resource-constrained drones to store a copy of the full blockchain. Drones can address, detect, recognize, and control disease infection outbreaks, which will lead to faster responses of illness containment such as malaria, COVID-19, as s shown in Table.2. In China, several advanced emerging technologies are viable technologies to mitigate the outbreak and spread of the COVID-19. For securing the internet of drones, the authors [62] addressed the intelligent technique called convolution neural network and compared it with the drone's autonomous internet. The findings showed better security performance of the proposed technique.

Furthermore, [63]introduced access control four-state models to avoid malicious nodes, provide high-security levels, and improve IoT devices' energy efficiency.

Table.2 Drone technology for disease prevention and medical supplies

| Ref. | Highlighted | Drone function | Blockchain function | Disease |
|---|---|---|---|---|
| [60] (2017) | It provides flexible and low cost to map water bodies in order to detect mosquito and disease elimination. | Sensing capability | | malaria |
| [64] (2014) | Tracking global burden of malaria by drone technology to disseminate larvicide | Sensing capability | | malaria |
| [65] (2014) | exploring drone for mapping environmental risks of zoonotic malaria | Sensing capability | | malaria |
| [66] (2018) | Drones for zoonosis controlling infectious diseases among animal populations. | Sensing capability | | malaria |
| [67] (2017) | Drone is used for drug delivery in quarantine areas undergoing epidemic outbreaks. | Transportation | | Drug delivery |
| [68] (2016) | Drone technology is used for transporting laboratory samples to the diagnosis of HIV in Malawi. | Transportation | | HIV |
| [4] (2020) | Blockchain for collaboration of multi-robot to fight COVID-19 | Transportation Monitoring | Make Consensus Decentralization Managing and control | COVID-19 |
| Current work | Blockchain for multi-drone decentralization to combat the COVID-19 outbreak | Transportation Detection Monitoring | Zero touches remote Decentralization Make consensus Drone traffic control | COVID-19 |

Blockchain technology represents a critical role in managing and controlling multi-drone efficiently and secures and ensuring accuracy. Moreover, blockchain helps to secure multi-drone via avoiding hackers and malicious, and controlling the multi-drone to maintain track (no crash, no harm, no injuring) during the performance of their tasks. Therefore, blockchain is working here as a "meta controller". The importance of using blockchain is to update drones' real-time location during delivering the medical supplies, monitoring the people who break the rules, and separating people gathering by producing speaking in quarantine areas. To avoid collisions of drones, data on the blockchain will be public, and therefore, each drone can access the locations of other drones, as shown in fig.2. Giving drones a unique identity on the blockchain while delivering medical supplies, IoT devices at home, and hospitals give drones trusted to access secure locations. The authors of [69] discussed how software and hardware setup of Ethereum to secure exchange information among multi-drone during performing the tasks. Once permission is confirmed, the doors and windows will open automatically, and the drone will deliver goods and complete door to door delivery task.

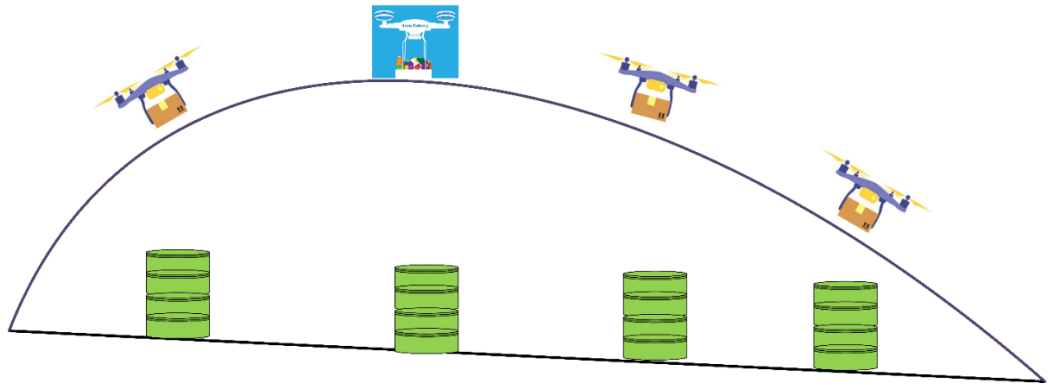

Fig.1 Drone and blockchain for traffic management

The network architecture of the blockchain-based multi-drone system is shown in fig.3. Fig.3 consists of a multi-drone, blockchain decentralized network, and control system. Based on blockchain, multi-drone can form a drone cluster to perform complex tasks, interact, real-time analysis, and process to facilitate decision making. Multi-drone has carried IoT devices. Furthermore, multi-drone requires communicating with each other and sharing collected data, flight status, speed, and location to combat COVID-19 collaboratively. Therefore, blockchain uses to collect data from drones and responses based on controlling and data storing in order to ensure the stability of sharing data, improve collaboration, and decentralize network to avoid malfunctioning and network partitioning.

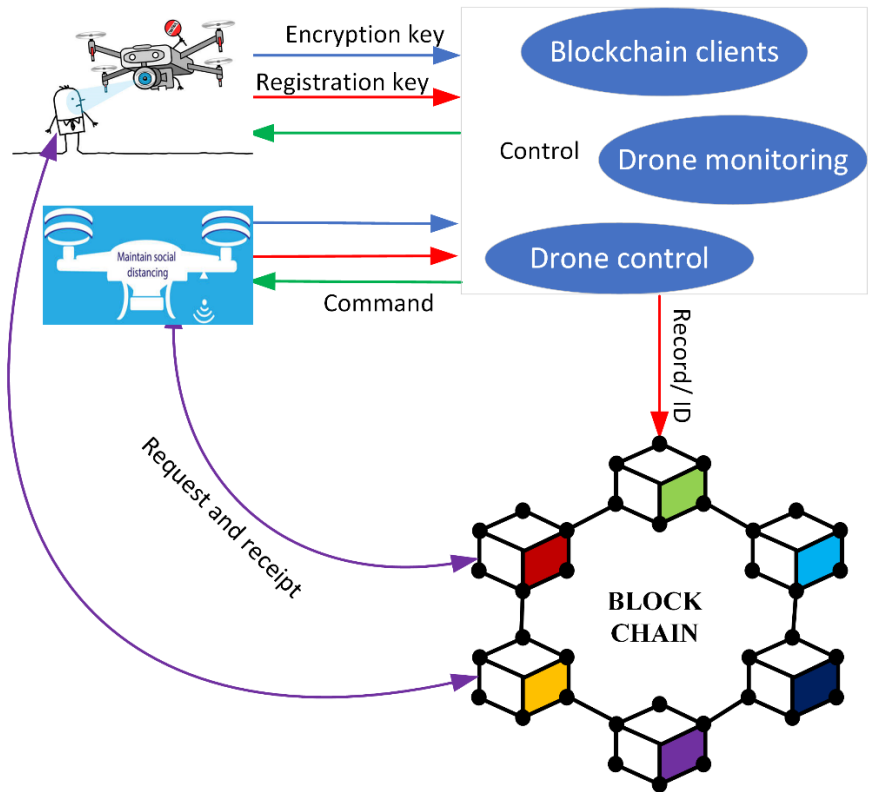

Fig.3 DroneChain communication architecture

a) Blockchain

Blockchain is promising in many domains of real applications of industrial and academics. Blockchain technology is essential to be used in the robotic domains because it can establish a reliable multi-robot network with a high-security level and reach overall consensus robots in a decentralized fashion. It stores data and transfers between robots in the network as transactions. Blockchain features ensure that adding a block to blockchain has more secure. The authors of [70] discussed the consensus algorithms including Proof of Work (PoW), Proof of Stake (PoS), Delegated Proof of Stake (DPoS), Practical Byzantine Fault Tolerance (PBFT), Proof of Capacity (PoC), Proof of Burn (PoB), etc. Blockchain types and characteristics are shown in table.3.

Table. 3 Blockchain types and features

| Blockchain types | Characteristics | | | | | |
|---|---|---|---|---|---|---|
| | Efficiency | Transaction | Read/ Write | Transparency | Scalability | Consensus |
| Private | High | Short | Permissioned | High | High | PBFT, RAFT |
| Public | Low | High | permissionless | Low | High | PoS, DPoS |
| Consortium | High | Short | Permissioned | High | Low | BPFT |

The combination of blockchain technology and multi-drone supports drones robustness, ease of scaling, operation security, flexibility, autonomy, and even profitability. To prevent and avoid collisions of robots, secure robot operations, and ensure robots stay on track. Blockchain and multi-drone combination enhances task automation, decision making, task formulation, authentication, and action validation.

Blockchain is the critical solution for distributed decision making of multi-agent collaboration. Simultaneously, the development of smart contract in blockchain technology for the collaboration of multi-drone can be built for formulating the proposed actions in bytecode. Then, multi-drone is going to collaborate efficiently with each other and vote for adaptive and effective action. For action validation, drones can monitor each other (i.e., locations, states, energy, and actions, etc.). If the drone sends incorrect data, the sharding technique can be applied to solve the validator issues [71]. Therefore, the consensus for the wrong action will be avoided. To make the task autonomously, blockchain distributes consensus for dispatching, executing and assigning tasks during multi-drone collaboration. Therefore, Blockchain can handle drone failures, attacks, and hijack efficiently. Moreover, blockchain can detect drone fails before it occurs and produces an alert in this regard. In this case, drone fail does not impact other drone tasks due to blockchain decentralized features.

**b)** DroneChain

DroneChain is a blockchain that is designed for drone operation to perform complex tasks. It is connected to drones to allow events storing among them inside DroneChain. The main characteristics of Dronechain are easy verification, self-amending & use of consensus algorithms. Furthermore, it supports accessibility, decentralization, auditability, non-reputation, irreversibility, secure data, and trustworthy [72]. DroneChain architecture is shown in Fig.2. Fig.2 is including drones, a decentralized blockchain network, and a control unit. For combating the COVID-19 outbreak, many drones are required for different purposes and needs, such as monitoring, detecting, and E2E delivery. Each drone carries different devices according to its activity in combating COVID-19. Decentralized blockchain network uses for data validation (i.e., data collected from drones, control unit actions, and integrity protection). Therefore, the blockchain decentralized network helps decentralize multi-drone collaboration in the quarantine area via data gathering from drones, control unit responses, data storing and sharing, and decision-making in real-time. Control unit requires decentralized multi-drone collaboration to send gathered data by drone, make commends, and update the event's status to drones to change their behaviors. Furthermore, it can help in sending original and hashed data to the blockchain network.

## IV. Proposed framework solution

Regarding combat the COVID-19 outbreak, the integration of blockchain technology and multi-drone collaboration makes robots ease of scaling, control, and robust against failure. To avoid the multi-drone collision, secure robot operations, and ensure robots stay on track in the quarantine area, blockchain can efficiently manage the robot traffic, support multi-drone collaboration. Blockchain makes data public for each drone, and therefore each robot can reach other drones data in high authentication and accuracy. For combating COVID-19, we focus on discussing the efficient solutions offered by using blockchain technology for identified decentralized multi-drone collaboration challenges. We discuss how blockchain technology will control and manage decentralized multi-drone collaboration to combat COVID-19. Then, we use cases in real-world applications of multi-drone collaboration to monitor and detect infected people in the quarantine areas and E2E delivery system.

Blockchain helps multi-drone collaboration to fight COVID-19. The drones need to reach a common goal based on their joint planning, and distributing their decision-making during planning is challenged. Therefore, flexible and autonomous solutions for multi-drone distributed decision-making require to tackle the challenges. Blockchain technology represents the critical solution for ensuring that all drones are connected in a decentralized fashion and share an identical world view. For instance, in every sub-task performed, each drone requires the agreement of other drones in order to act accordingly. However, with using blockchain, the information of action is in the block and available for public drones in the group.

Fig.4 is shown the combination of consensus algorithm and sharding technique for improving blockchain scalability and developing decentralized blockchain to decentralized multi-drone collaboration in combating COVID-19 in the quarantine areas. The consensus algorithm supports decentralized multi-drone collaboration in blockchain network for combating COVID-19. At the same time, the sharding technique improves blockchain scalability.

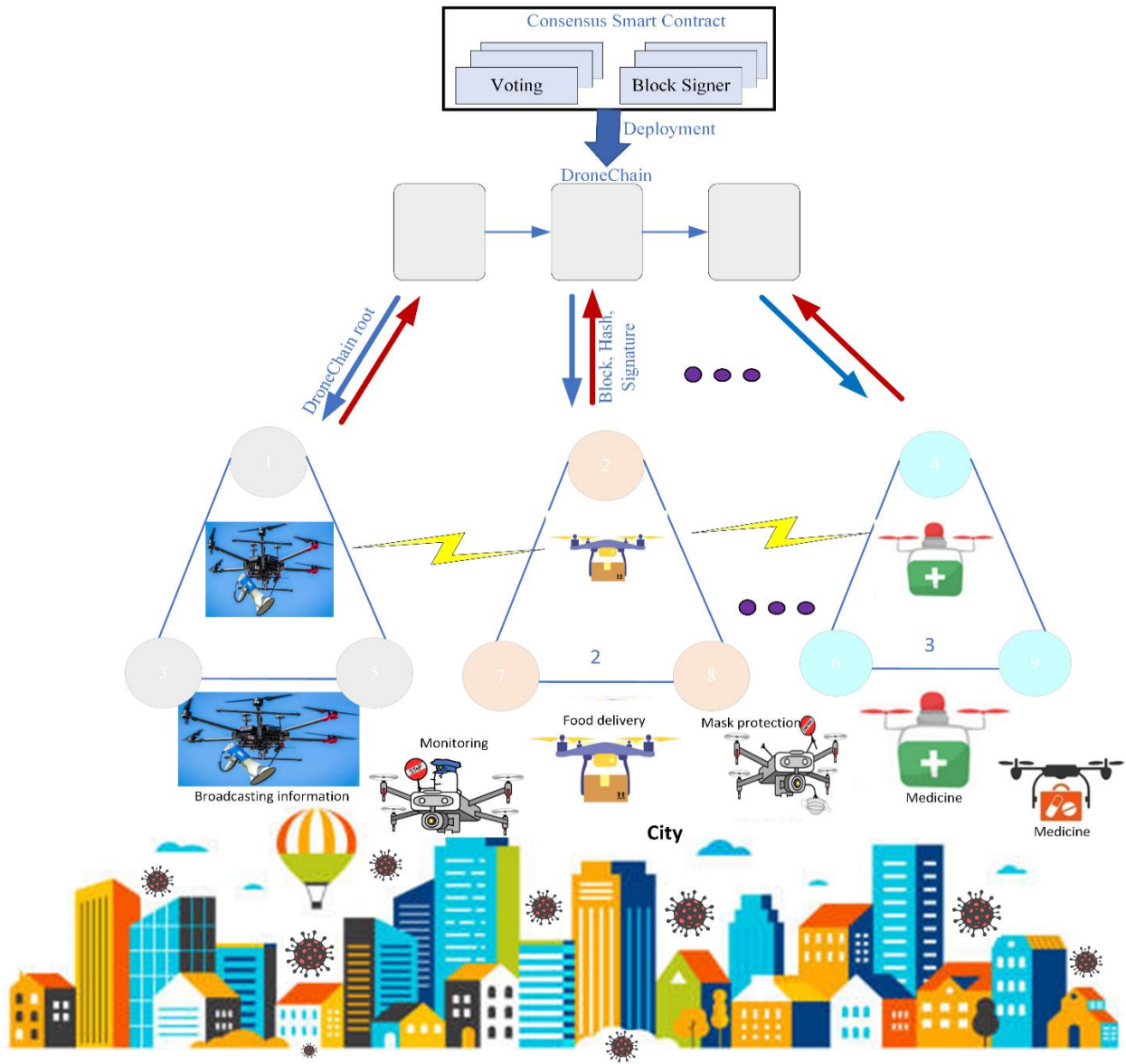

Fig.4 Decentralized and proposed technique for multi-drone collaboration to fight COVID-19

A. The consensus algorithm for supporting decentralized multi-drone collaboration

Consensus algorithms help multi-drone collaboration decentralized blockchain network to reach an agreement. Table.4 illustrates several famous consensuses algorithms (i.e., PoW, PoS, and PBFT), benefits, and limitations that are used for blockchain development. Therefore, each drone in a decentralized blockchain network trusts all robots. The authors of [73] introduced the consensus algorithm for managing decentralized multi-drone collaboration. In [70], the authors discussed and highlighted the consensus algorithms with details. However, applying the consensus algorithms in blockchain can provide scalability, decentralized, and secure multi-drone agreement [74]. To allow a multi-drone to reach the consensus, the fault-tolerant consensus is made a trade-off within a triangle formed and required latency.

Table.4 Consensus algorithms benefit and applications

| Consensus | Benefits | Limitation |
|---|---|---|
| PoW | Secure and decentralization | High latency and high resource |

| PoS | Secure and low resource | Low participation |
| PBFT | Low latency and high throughput | Secure and Low scalability |
| DPoS | Secure, low latency, and high throughput | Producer evil |

A. Blockchain scalability for decentralized multi-drone collaboration

A significant effort is recently given to the development of decentralization, scalable, and secure blockchain, which is designed in Ethereum 2.0 [75]. Therefore, computational resources are not needed for mining [76]. PoW is used for shard scalability and a secure blockchain network. Shard synchronization and executing the contract from every single shard must be deployed in the blockchain network's smart contract. Shard synchronization allows the flow of data between shards. Therefore, the sharding technique improves the blockchain scalability and improves accommodates fault tolerance. Moreover, the combination of the dynamic sharding technique and the consensus algorithms in location, community (membership), task types, population, etc., support detecting malfunctioning drone and network partitioning. Membership [28] and DAG [28, 77] techniques can be integrated with sharding techniques. The integration will establish multi-drone collaboration based on dynamic sharding the consensus algorithm on blockchain technology. It would result in realistic collaboration with avoiding collision, network partitioning, and failure tolerance.

In the case of scalability, blockchain can support multi-drone operators and maintenance in distributed decision-making processes. Because of all planning actions, agreements are based on the transactions stored in the blockchain. Therefore, it is not required to train new drones joining the team to perform common goals or if anyone has failed. New drones can download the ledger that contains all of the agreement history contained the previous knowledge, store it, and then start planning action accordingly. In the case of a drone malfunctioning, it can be replaced by another drone. If any element/device equipped drone malfunctioning, then data of that element/device can be borrowed from his neighbors. Blockchain becomes critical because of supporting decentralized applications for sharing data in real-time among multi-drone collaboration to combat the COVID-19 outbreak. With blockchain, multi-drone collaboration can play a crucial role in monitoring, detecting, managing, and controlling public health emergencies during COVID-19 and future pandemics.

B. Application of proposed framework (Some work required here)

Multi-drone plays a vital role in delivering medical supplies and goods to people quarantined at homes or hospitals due to the COVID-19 outbreak. Therefore, the interaction between humans is minimized because of drones function (i.e., autonomous spray disinfectant, delivering medical supplies and goods, controlling infection by sperate stuff from patients, keep health workers out of the quarantine area, identify the infected case, detected person who breaks quarantined restrictions rules). Hence, drones will satisfy the patients quality of experience in the quarantine due to getting the staff's care and healthier people without risk. The collision risk increases gradually due to the number of drones is growing increasingly in the quarantine area for delivering things, monitoring, detecting, and controlling.

Blockchain helps update each drone's locations on the distributed ledger, while each drone can reach all drone locations updated in the smart contract to avoid collisions, as shown in fig.5. For instance, all drones deployed for a different purpose in the quarantine area (i.e., detecting and identifying infected cases, broadcasting news, monitoring people who break the quarantine rules, delivering the medical supplies, delivering goods to people in the quarantine area, and automating spray disinfectant) update their location, speed, the status of the battery, delivery time and date, and resources to blockchain network autonomously and in real-time. Each drone can access other drones locations due to blockchain making data public for all drones, and therefore, drones can avoid collisions. Therefore, blockchain represents the critical solution for drones traffic management efficiently with high accuracy. Moreover, blockchain helps secure drones operation by ensuring that all drones in the network stay on track (no harm, no crash, no injuring).

1. Drones for door-to-door delivery

In this scenario, the door-to-door delivery system is required for combating COVID-19 via drop the delivered package to the customer. Furthermore, pick up and delivering a medical sample requires a fast delivery system. Drones represent the critical technology for transportation during COVID-19. Drones can combat the COVID-19 outbreak by delivering goods to the patients in a smart hospital or to the people who are quarantined restricted in smart homes without human interaction. However, opening and closing doors require smart devices (IoT devices), which may be considered in smart cities. Therefore, a combination of multi-drone collaboration and IoT devices play a vital role in delivering orders/goods in the quarantine area to combat COVID-19. IoT devices enable the window/door to interact with drones to drop a package to the quarantined person either in a hospital or at home.

Here, blockchain is used to store drones updated data to ensure the door/window open for a certain drone. The drone reached the near arrived house in the quarantine area, blockchain updated the location, and shared it with a specific home's window/door. Therefore, IoT devices in the supposed house's window/door enable the window/door to identify and verify the blockchain to open and let drones enter the house. As soon as a specific drone for delivering a particular task reaches the receiver location, then the smart house's door/window will be open due to the automatic. Then, drones deliver goods to the patient table or quarantine restricted people at a smart home. While the drones go out, the door/window closed and then go to perform another task. Furthermore, if any documents are required from anyone quarantine area, the drone can perform the task of scanning identity card or any documents and send them to the required office, as shown in Fig.5-B.

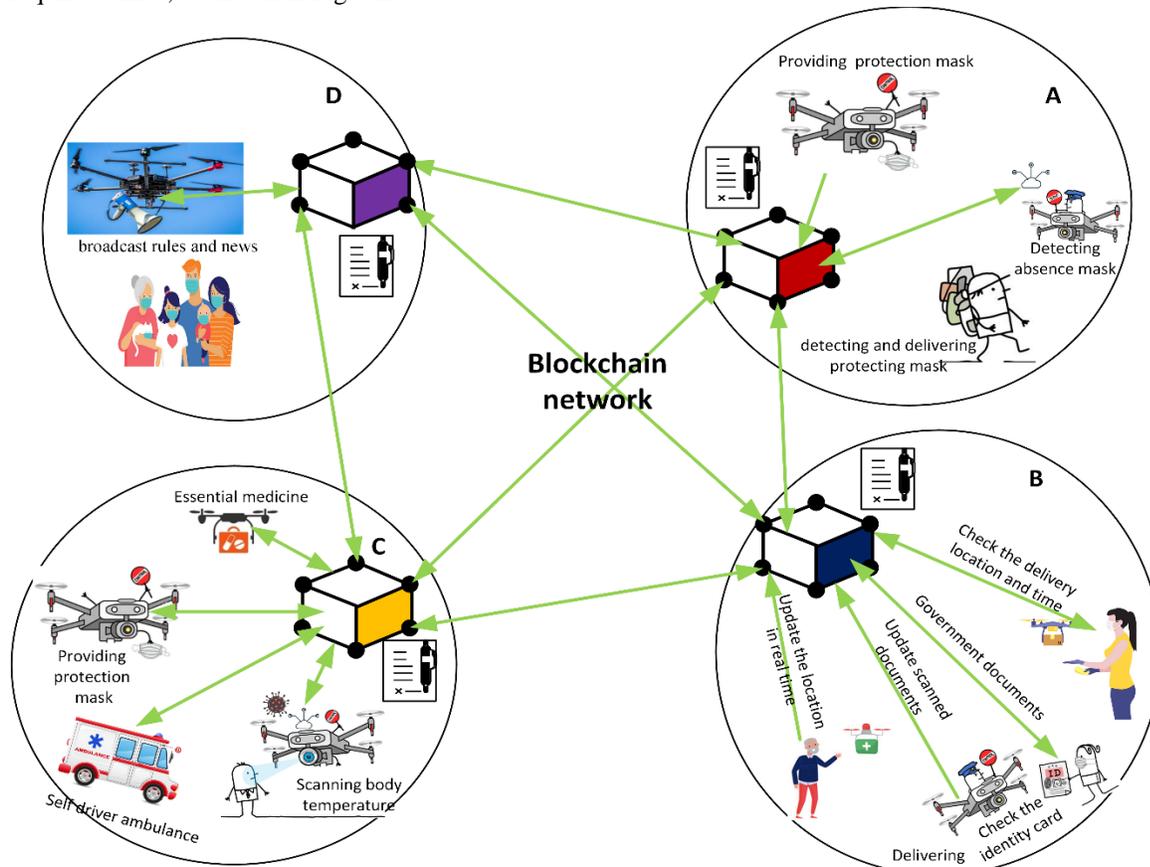

Fig.5 Application of blockchain for decentralized multi-drone collaboration

2.  Broadcasting Practicing social distance

In this scenario, two cases for using a drone are shown in fig.5-D. The first case is drone monitoring people gathering and practicing social distance. In the second case, the drone is used to broadcast news about the COVID-19 danger when people gather even if they wear masks. Drone can stop people from gathering and monitoring the social distance between people during their door or outdoor activities. The importance of blockchain here is to produce alerts to police if people do not follow the drone guiding.

3.  Testing body temperature

In this scenario, the drone detected an infected person of COVID-19 with an absence mask. The process of identifying infected cases can be done in the drone. Drone update the location, case level, and requirement to the smart contract in the blockchain. Then, blockchain will produce alerts to the ambulance and other drones such as spray disinfection drones, drones providing protection masks, etc. The notified drones and ambulance will access the smart contract and read all of the detection cases' requirements, as shown in fig.5-C. Fig.5-C shows how few drones are collaborating to combat COVID-19 in a decentralized fashion achieved by blockchain technology.

4. Absence mask detection

In this scenario, the drone is used to detect absence masks, while the absence mask process is done in a drone using deep learning for face recognition, as shown in fig.5-A. Furthermore, another drone is used for mask delivery if detecting a person is not responding to the drone and wear a mask or may not have a mask based on notification from the blockchain. On the other hand, blockchain will alert police to take necessary action against people who break quarantine rules.

## V. Use cases of the proposed framework

### A. Multi-drone collaboration for E2E delivery system

Drones are a promising technology in delivering goods to people in the quarantine area with support zero-touch remote. Therefore, drones support combating COVID-19 by transporting medicine, goods. We focus on managing and controlling multi-drone during the E2E delivery system using advanced technology to perform tasks effectively and efficiently. Drone delivery time, reliability, and network connectivity are required for 5G to implement the use cases. 5G connectivity is preferable due to high quality, connectivity, and communication security among multi-drone collaboration networks to combating COVID-19. Multi-drone collaboration needs for security attack in order to control spoof attack vectors. The attack vectors include data monitoring and data acquisition. For avoiding attack vectors, blockchain technology requires multi-drone collaboration in the E2E delivery system. Furthermore, 5G offers reliable data, access data quickly, high data availability for the combination of multi-drone and blockchain.

The proposed framework for the E2E delivery system of the multi-drone to combat COVID-19 is shown in Fig.6. It consists of three layers, i.e., received Rx, blockchain, and donate Dx layers. In the Received layer, people in the quarantine area request/order food and medicine to prepare in the smart contract (order, person location, denote location, etc). This request is passed through the blockchain network the send to the donate layer to prepare the goods. The request will be prepared with encrypted data in the donated layer, including drone ID, path, receiver location, and delivery time. The smart contract executes and verifies the condition and status between the donated layer and receiver layer. Then, the verified block will be sent to the blockchain. All data stored in the blockchain is secure and reliable. The required drone will accept to perform and deliver the order from accessing smart contract and read the receiver's location and denote. Then, the drone ID updates to the smart contract. Both receiver and donate layer update the drone ID trajectory and time. Drone maintains to update its location simultaneously to blockchain in order to inform both donate and receiver the status and condition. Finally, drone update to the blockchain the order has been successfully delivering via updating returning and dispatching. Fig.7 shows the procedure and algorithm of the E2E delivery system with blockchain technology in decentralized multi-drone collaboration during combating COVID-19 in the quarantine area.

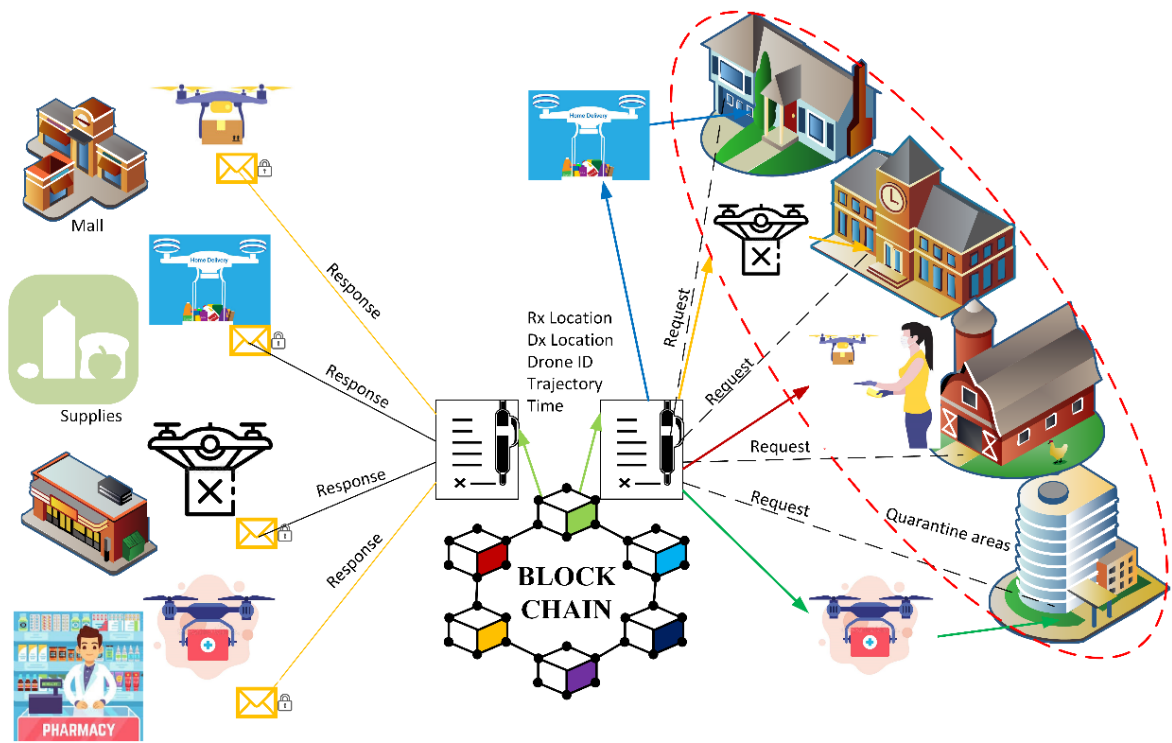

Fig.6 E2E delivery system using multi-drone collaboration and blockchain technology

During the flight, drones must be coordinated to keep their trajectory to reach the goal and avoid a collision. Therefore, blockchain is working as a meta controller to control the drone's flight and avoid a collision. Blockchain is a decentralized technology that can coordinate the drones flight allocation in a decentralized fashion. So, drones allocation/ position always updates and is stored in the ledger and managed in peer-to-peer networks. Before drones mission in space, each drone will add a block to the blockchain requesting fight trajectory while adding the initial and destination position. Then, blockchain will ensure the flight trajectory and coordinate drone missions with others already available in space.

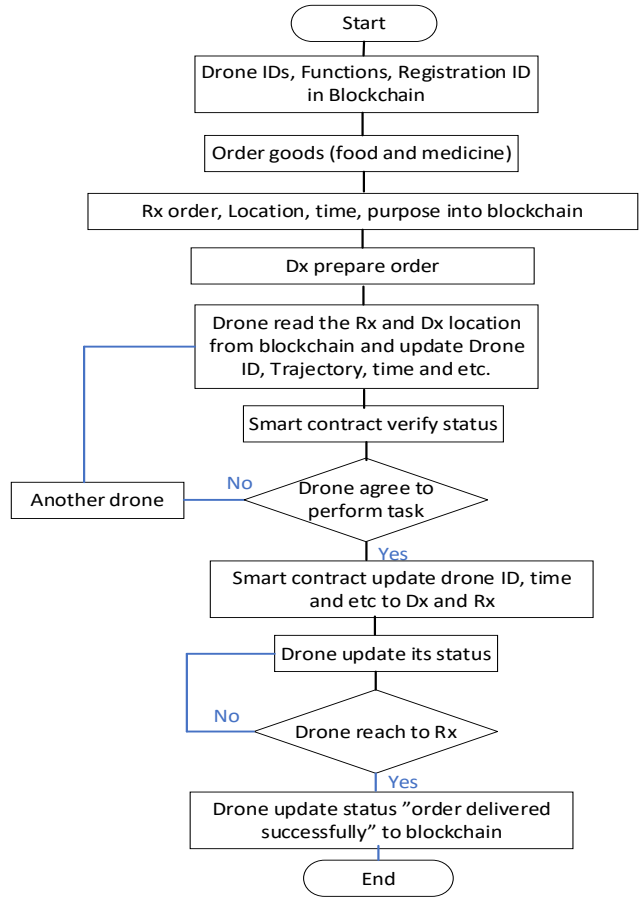

Fig.7 Flowchart of outdoor E2E delivery system

### B. Multi-drone collaboration for monitoring and detecting COVID-19

Drones are equipped with smart IoT devices, high-resolution cameras, computers, and thermal cameras for different purposes and needs. For instance, the camera can detect people who do not wear masks and do not practice social distance during COVID -19 outbreak and future pandemics. At the same time, a thermal camera can detect the infected person of COVID-19 earlier from a long distance. Also, computers can be used to analyze captured date from drone equipment locally as an edge intelligence within AI techniques. Furthermore, drones are used for spraying disinfection and delivering goods, foods, and medicine during pandemics such as COVID-19.

Blockchain works as "Meta controller". Blockchain is a crucial technology for deploying multi-drones for different purposes to quarantine area with avoiding collision. The blockchain stores and shares information to all drones in the blockchain network in a decentralized fashion. A smart contract is then used to control multi-drone to perform their tasks efficiently and effectively without collision. In the case of combating the COVID-19 outbreak, the smart contract also helps multi-drones to interact with each other to perform the shared task and make a decision in real-time. For instance, if a drone detects one infected man, then the status will be updated to the blockchain, and each drone reads the block and takes action according to its duty. Announce drone comes and guides people, disinfection drone joins the action by spraying disinfection around the detected man, and drone with high-resolution camera monitoring area as shown in Fig.8. An intelligent algorithm is implemented in the smart contract by detecting an infected person, practicing social distance, or absence mask. Smart contracts can identify the needed action according to captured data received from a drone, then send action to drones for changing behavior to perform required tasks accordingly.

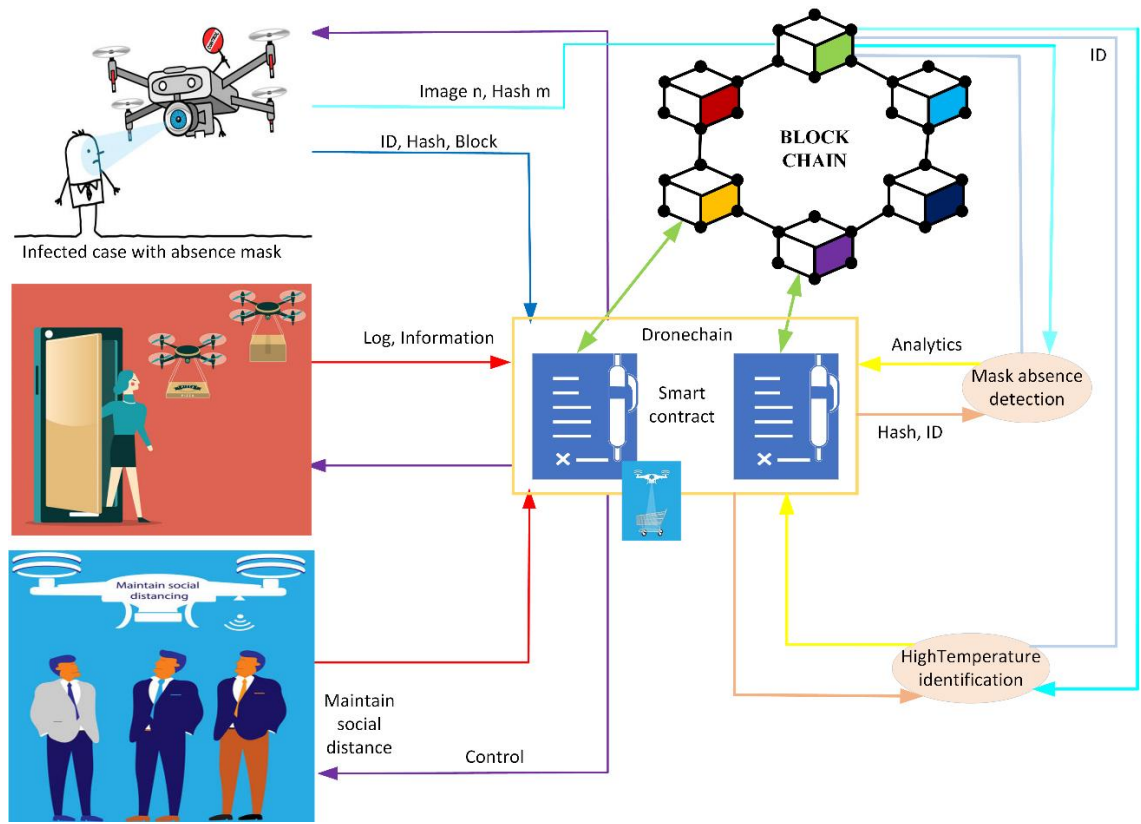

Fig.8 Multi-drone collaboration for monitoring and detecting COVID-19

In the absence mask case, the drone can detect a person with an absent mask via a high-resolution camera equipped in the drone body. Furthermore, the drone is considered an edge intelligent to analyze image locally with the help of AI. Then drone produces an alert to the detected person immediately. The drone will update the information into the blockchain and immediately register (drone ID, person movement, location, people around, etc.) as shown in Fig.8. If the drone is still updating alert, the mask detection unit will update action to the drone chain, and another drone will identify the person's location and deliver mask to him immediately as $0^{th}$ responded for saving people life and avoiding COVID-19 outbreak. Furthermore, if people gathering detected even if they wear a mask, a drone will produce an alert for practicing social distance and avoiding interaction.

In case the infected person in a critical case, the doctor can access particular patient status in blockchain and recommend suitable medicine to smart contracts. Pharmacist can access smart contracts and then prepare the required medicine suggested by the doctor. Then, another medical delivery drone is used to deliver medicine from the Pharmacy to the patient table in the smart hospital case, including explaining how to use the medicine. Therefore, the interaction between humans is eliminated, and therefore, drone collaboration plays a critical role in fighting the COVID-19 outbreak efficiently in a high-security manner.

If a high body temperature is detected drone, the drone will update the information (block, hash, and drone ID) to dronechain. It will then be registered in dronechain (drone ID, Location, environment status) as shown in fig.8. All drones log information and read status from dronechain to respond immediately to perform their task collaboratively. The disinfection drone will spray disinfection in the detect person environment. The medical delivery drone can deliver the required medicine. Many drones are deployed to find if COVID-19 infects someone else. Furthermore, a self-driver car joins the action to carry the infected person to the hospital. Based on the smart contract in dronechain, all drones respond to action and change their behavior according to the need in a decentralized fashion without collision. The procedures are summaries in the following algorithm, as shown in Fig.9.

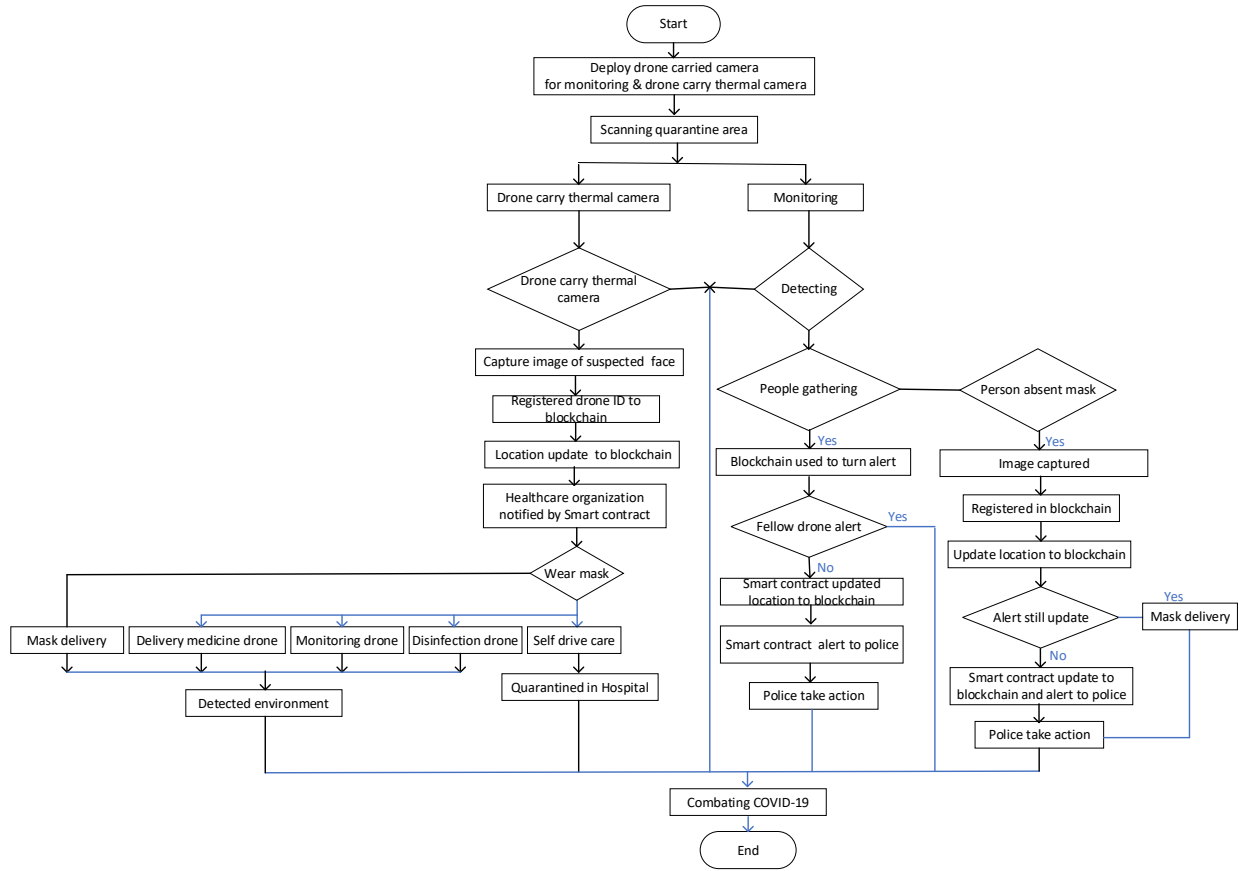

Fig.9 Flowchart of monitoring multi-drone collaboration to combat COVID-19

## VI. Discussion

Recently, drone technology is a promising solution in avoiding humans interaction, especially during the COVID-19 outbreak and future pandemics, by monitoring, detecting (mask absence, high body temperature), spray disinfection, delivering goods, and broadcasting information, etc.

### A. Challenges

This paper defines the proposed conceptual framework solutions for decentralized multi-drone collaboration using blockchain technology in order to combat the COVID-19 outbreak and future pandemics. The limitations and challenges of the proposed framework solutions include lifecycles of blockchain technology and guarantee the new joined drone to decentralized blockchain network provides its real data quality, not a hacker. Gathered data from smart IoT devices may also impact the resources of drones.

For thermal scanning in real-time and performing simulation-based experience, residents have to be considered for scanning. Small drones can fly to perform the resident indoors (inside the building) if they cannot go to balconies for scanning. However, the scanned image analysis will take time due to the small drone's capacity to gather data without interruption is an open issue and should be taken for future consideration.

### B. Opportunities

We observe opportunities related to combat COVID-19 outbreaks by monitoring, detecting, and E2E delivery systems for quarantine areas. In the smart cities, combating the COVID-19 outbreak and future pandemics is more manageable than others. For smart home and smart hospitals, IoT devices placed in windows or doors should be nodes in the blockchain network. The multi-drone can then collaborate with smart home and smart hospital infrastructure (smart IoT devices) based on the blockchain network. Therefore, the combination of multi-drone and smart IoT devices in a smart environment (smart hospital, smart home) can allow drones to deliver goods to people inside the hospital or room. Furthermore, blockchain provides data to multi-drone and then validates the data stamps send from multi-drone in the same environment.

For the scanning and recording of residents in the quarantine area, multi- drones can be deployed for performing these tasks during the COVID-19 pandemic. Thus, this task is essential to record all and update it to match scanned people in a particular area. Furthermore, developing an integrated medical service, gathering data, fast monitoring, and deploying are needed. Collecting data from large areas for a long time is an open issue due to drone battery life limitations.

In the case of E2E delivery in the smart city, the window could have a sensor that can sense the drone when reaching the drone. In this case, the window sensor should be one node of the blockchain network to update the drone location on time. The drone can access the room via the window and deliver things to the room with zero touches (no one touches the things deliver by drone).

## VII. Conclusion

The integration of multi-drone collaboration and blockchain improves the way we care for sick people and infectious disease outbreaks in the quarantine area. In blockchain, decentralization is promoted to improve multi-drone performance and reduce time spent on performing a specific task. The multi-drone task is to combat the COVID-19 outbreak by delivering medical supplies, goods, and monitoring with reporting people those breaking rules of the quarantine. Therefore, blockchain adapted to support many events very fast and deal with many interacting drones in a decentralized fashion. For improving scalability and network patriation, a consensus algorithm is efficient for improving the network patriation of drones, while, sharding scheme uses to increase blockchain scalability. Our proposed conceptual framework solutions provide future potential experiments and domains that will develop blockchain-based decentralized multi-drone collaboration to fight the COVID-19 outbreak and future pandemics. Moreover, this conceptual framework will motivate industrial and academic researchers to pay more attention and efforts to implement the used case experimentally. Furthermore, it encourages researchers to explore the new use cases, in which the combination of emerged and advanced technologies can be used together (i.e., robots, drones, AI, blockchain, digital twins, and smart IoT technologies) to help fight the future pandemics; like COVID-19.

Funding: This project has received funding from the European Union's Horizon 2020 research and innovation programme under the Marie Skłodowska-Curie grant agreement No. 847577; and a research grant from Science Foundation Ireland (SFI) under Grant Number 16/RC/3918 (Ireland's European Structural and Investment Funds Programmes and the European Regional Development Fund 2014-2020)